# SOME INTEGRATED PROPERTIES OF GALACTIC GLOBULAR CLUSTERS

Sidney van den Bergh

Dominion Astrophysical Observatory

5071 West Saanich Road

Victoria, British Columbia, V8X 4M6

Canada

Electronic mail: vandenbergh@dao.nrc.ca

**ABSTRACT**

The luminosities of globular clusters are found to correlate with their half-light radii. The most luminous clusters have radii $r_h \sim 3$ pc. Mean cluster luminosities are $<M_V> = -6.64 \pm 0.26$ for $r_h < 2.0$, $<M_V> = -7.44 \pm 0.20$ for $2.0 \leq r_h < 4.0$ pc, and $<M_V> = -6.57 \pm 0.21$ for clusters with $r_h \geq 4.0$ pc. An even fainter value $<M_V> = -5.85 \pm 0.36$ is found for large clusters with $r_h \geq 8.0$ pc. These results possibly weaken confidence in the conclusion that the peak of the globular cluster luminosity distribution is a universal standard candle.

In the outer Galactic halo globular clusters with red horizontal branches are fainter by about a factor of ten than are clusters with blue and intermediate color horizontal branches. Among clusters with $R_{GC} > 10$ kpc there appears to be a clear dichotomy between normal clusters (which all have [Fe/H] < -1.2) and the anomalous relatively metal-rich clusters Pal 1, Pal 12 and Ter 7, which are both unusually faint ($M_V > -5$) and relatively metal-rich ([Fe/H] > -1.0). This suggests that these relatively metal-rich halo clusters may have had an unusual evolutionary history.

**1.    INTRODUCTION**

The availability of CCD detectors has, in recent years, produced explosive growth in both the quantity and quality of data on globular clusters. As a result of this a vastly expanded data base (Harris 1996a) is now available to explore the relationships between the integrated properties of globular clusters. These new data are already beginning to place significant constraints (cf. van den Bergh 1995a) on evolutionary scenarios for our Milky Way System.

The organization of the present paper is as follows: In § 2 evidence is discussed which suggests that the luminosity distribution of globular clusters depends on their half-light radii $r_h$. Subsequently, it is shown in § 3 that the integrated luminosity $M_V$ of clusters in the outer halo depends on the stellar population gradient along the cluster horizontal branch. Finally, § 4 draws attention to three anomalous faint, and relatively metal-rich, globular clusters that are located in the outer Galactic halo.

**2.    RELATION BETWEEN LUMINOSITY AND RADIUS**

A number of recent papers (Blakeslee & Tonrey 1996, Harris 1996b, and Sandage & Tammann 1995) have examined the question whether $M_V^o$, the peak of the globular cluster luminosity function, is a universal standard candle. This question is of importance because $M_V^o$ has recently been used (e.g. Jacoby et al. 1992, Whitmore 1995) to calibrate the extragalactic distance scale. A strong argument in favor of the view that $M_V^o$ might be a standard candle, that is not affected by environmental factors, was provided by Armandroff (1989). He showed that metal-poor ([Fe/H] $\leq$ -0.8) Galactic halo clusters and relatively metal-rich ([Fe/H] > -0.8) "disk" globulars (which must have formed in rather different environments) have similar values of $M_V^o$. Specifically, Armandroff found that $M_V^o$ = -7.50 $\pm$ 0.14 for 76 halo clusters, compared to $M_V^o$ = -7.52 $\pm$ 0.28 for 20 disk globulars. On the other hand Blakeslee & Tonrey (1996) have recently presented data which seem to favor the view that $M_V^o$ becomes fainter as the local density of galaxies increases.

Some evidence will now be examined which appears to suggest that the luminosity distribution of globular clusters depends on cluster half-light radius $r_h$, which is known to correlate with Galactocentric distance (van den Bergh & Morbey 1984, van den Bergh, Morbey & Pazder 1991, Djorgovski & Meylan 1994) and with perigalactic distance (van den Bergh 1995b).

Fig. 1 shows a plot of the absolute magnitudes $M_V$ of individual Galactic globular clusters versus their half-light radii $r_h$. This plot is based on a compilation of globular cluster distances and apparent radii (complete to 1996 April) which was kindly provided by Harris (1996a). This plot shows the following: (a) the dispersion in log $r_h$ is relatively small for luminous clusters with $M_V < M_V^o$. However, this dispersion increases rapidly towards fainter values of $M_V$. (b) Globular clusters with $2.0 \leq r_h$ (pc) $< 4.0$ pc have $<M_V> = -7.44 \pm 0.20$ which makes them more luminous than either small cluster with $r_h < 2.0$ pc, which have $<M_V> = -6.64 \pm 0.26$, or larger clusters with $r_h \geq 4.0$ pc, for which $<M_V> = -6.57 \pm 0.21$. The faintest clusters are those with $r_h \geq 8.0$ pc for which $<M_V> = -5.85 \pm 0.36$. [A notable exception is the very distant globular cluster NGC 2419 with $M_V = -9.5$ and $r_h = 19$ pc.] Since the largest clusters generally occur in the outer halo of the Galaxy this is equivalent to the conclusion that clusters in the outer halo have below-average luminosities (see Fig. 5 of van den Bergh 1995b) and Kavelaars & Hanes (1996). A Kolmogorov-Smirnov (K-S) test shows that this probability that globular clusters with $r_h < 8.0$ pc, and those with $r_h \geq 8.0$ pc, were drawn from a parent population with the same luminosity distribution is $< 2\%$. By the same token K-S tests show a probability of only 4% that clusters with $r_h < 2.0$ pc, and those with $2.0 \leq r_h < 4.0$ pc, were drawn from the same luminosity distribution of parent objects. Finally a K-S test gives $< 1\%$ probability that intermediate-size globulars with $2.0 \leq r_h < 4.0$ pc and large clusters with $r_h \geq 4.0$ pc were drawn from the same parent population of luminosities.

An error which produces too large a cluster distance will give a cluster luminosity that is too high and a cluster diameter that is too large. Observational errors will therefore tend to exaggerate the difference in luminosity between small and intermediate-size globulars. However, such errors would also make the true luminosity differences between clusters with $2.0 \leq r_h < 4.0$ pc, and those with $r_h \geq 4.0$ pc, even greater than the difference found above. (c) Inspection of Fig. 1 shows that the most metal-rich clusters ([Fe/H] > -1.0), which are plotted as open circles, tend to have below-average values of $r_h$. This is, no doubt, due to the fact that the smallest globular clusters occur almost exclusively close to the Galactic center (van den Bergh & Morbey 1984, van den Bergh, Morbey & Pazder 1991, Djorgovski & Meylan 1994), where the mean cluster metallicity is high.

Histograms of the luminosity distributions of globular clusters in four radius bins are shown in Figure 2. The differences between these distributions suggests that conditions prevailing at the time of cluster formation may have affected both cluster size and cluster luminosity. It is not immediately obvious how more recent environmental factors, such as disk/bulge shocks or evaporation of clusters (Murali & Weinberg 1996, Gnedin & Ostriker 1996) could account for the fact that the brightest globular clusters have $r_h \approx 3$ pc, whereas both larger <u>and</u> smaller clusters have fainter mean luminosities. Possibly two-body relaxation drives expansion and fast dissolution of compact clusters, whereas tides and gravitationally shocks might decimate the population of the largest clusters. In any case it appears prudent to assume that both different initial conditions, and differing environmental factors, might have affected the luminosity distribution of globular clusters in different galaxies. This suggests that one should exercise caution when using $M_V^o$, the peak of the globular cluster luminosity function, as a standard candle for calibrating the extragalactic distance scale.

## 3. CLUSTER LUMINOSITY AND HB GRADIENTS

Figures 3 and 4 show plots of $M_V$ versus $R_{GC}$ for globular clusters with red [$C \equiv (B-R)/(B + V + R) < -0.80$], and with blue and intermediate-color [$-0.80 \leq C \leq +1.00$] horizontal branches, respectively. Intercomparison of Figures 3 and 4 shows a significant difference between these two types of clusters in the outer halo ($R_{GC} > 10$ kpc) of the Galaxy, but no obvious difference in the inner halo ($R_{GC} \leq 10$ kpc). For clusters with red horizontal branches that are located in the outer halo $< M_V > = -4.82$, which is almost ten times less luminous than the value $< M_V > = -7.30$, that is found for outer halo clusters with blue and intermediate-color horizontal branches. A Kolmogorov-Smirnov test shows that there is only a 0.1% chance that the red and blue HB clusters with $R_{GC} > 10$ kpc were drawn from the same parent population if cluster luminosities.

At a given metallicity level globular clusters with red horizontal branches are believed to be younger than ones that have blue horizontal branches (Rood & Iben 1968, Rood 1973, but see Richer et al. (1996)). So the observed effect <u>might</u> be due to a decrease in the luminosity with which clusters are formed in the outer halo over time. Alternatively, this difference might be related to the fact that red horizontal branch clusters in the outer halo are, in the mean, more metal rich ($<$ [Fe/H] $>$ = -1.32) than are clusters with bluer horizontal branches ($<$ [Fe/H] $>$ = -1.70). Perhaps second generation globular clusters that formed in the outer Galactic halo were, on average, both less luminous <u>and</u> slightly more metal-rich than those formed earlier. However, the low metallicity ([Fe/H] = -1.69) of Ruprecht 106, which is the youngest known halo globular cluster (Kaluzny, Krzeminski & Mazur 1995), would appear to militate against such a simple scenario.

Intercomparison of Fig. 3 and Fig. 4 shows no significant dependence of the luminosity distribution of inner halo clusters with $R_{GC} \leq 10$ kpc on horizontal branch morphology. Furthermore, Fig. 4 shows no obvious difference between the luminosity distributions of clusters with blue and intermediate-color horizontal branches in the inner halo ($R_{GC} \leq 10$ kpc) and in the outer halo ($R_{GC} > 10$ kpc).

## 4. ANOMALOUS CLUSTERS

Figure 5 shows a plot of the absolute magnitude $M_V$ versus metallicity [Fe/H] for Galactic globular clusters with $R_{GC} \leq 10$ kpc. The data in this plot were taken from the recent compilation by Harris (1996a). The data, which are plotted in Fig. 5, cover a range of ~100 in metallicity. They show no evidence for a significant dependence of $M_V$ on [Fe/H].

Figure 6 exhibits a similar plot of $M_V$ versus [Fe/H] for outer halo clusters with $R_{GC} > 10$ kpc. This figure shows the following: (a) there is no significant evidence for a dependence of $M_V$ on [Fe/H] for clusters having a range of ~10 in metallicity. (b) A rather sharp cut-off in the metallicity distribution of outer halo clusters occurs at [Fe/H] $\simeq$ -1.2. (c) The three outer halo clusters Palomar 1, Palomar 12 and Terzan 7 are all relatively metal-rich ([Fe/H] > -1.0) and of below-average luminosity ($M_V > -5$). The location of Pal 1, Pal 12 and Ter 7 in Fig. 6 suggests that these clusters might have had an unusual evolutionary history. Richer et al. (1996) point out that Pal 12 and Ter 7 are 3-4 Gyr younger than other globular clusters of similar metallicity. It is noted in passing that Ter 7 is probably associated with the Sagittarius dwarf galaxy (Ibata, Gilmore & Irwin 1995).

## 5. CONCLUSIONS

A number of independent lines of evidence suggest that there are significant differences between globular clusters in the inner ($R_{GC} \leq 10$ kpc) and outer ($R_{GC} > 10$ kpc) regions of the Galactic halo. Red horizontal branch clusters in the outer halo are found to be an order of magnitude less luminous than are globulars with bluer horizontal branches. The low luminosity and relatively high metallicity of the outer halo clusters Pal 1, Pal 12 and Ter 7 suggests that these objects have had a different evolutionary history from that of ordinary metal-poor globular clusters in the outer Galactic halo. The dependence of the luminosity distribution of globular clusters on cluster radius raises some disturbing questions about the validity of the assumption that $M_V^o$, the peak of the cluster luminosity distribution, is a universal standard candle.

It is a pleasure to thank Bill Harris for providing me with an updated version of his globular cluster data base. I am also indebted to Don VandenBerg, Oleg Gnedin and to Martin Weinberg for thoughts on the interpretation of some of the data discussed above.

**TABLE 1.**   Absolute magnitudes and half-light radii derived from Harris (1996a)

| $r_h < 2.0$ pc | | | $2.0 \leq r_h < 4.0$ pc | | | $r_h \geq 4.0$ pc | | |
| --- | --- | --- | --- | --- | --- | --- | --- | --- |
| ID | $r_h$ (pc) | $M_V$ | ID | $r_h$ (pc) | $M_V$ | ID | $r_h$ (pc) | $M_V$ |
| N362 | 1.88 | -8.26 | N104 | 3.33 | -9.26 | N288 | 5.28 | -6.54 |
| Pal. 1 | 1.28 | -1.77 | N1261 | 3.32 | -7.68 | AM1 | 16.86 | -4.60 |
| Pal. 2 | 1.12 | -6.86 | N1904 | 2.84 | -7.73 | Eri. | 9.08: | -4.82 |
| N1851 | 1.77 | -8.26 | N2298 | 2.36 | -6.19 | N2419 | 19.19 | -9.48 |
| N2808 | 1.97 | -9.26 | E3 | 2.40 | -2.61 | Pyx. | 13.71: | -5.59 |
| N6093 | 1.59 | -7.85 | N3201 | 3.82 | -7.34 | Pal. 3 | 16.65 | -5.52 |
| N6287 | 1.79 | -7.06 | N4147 | 2.29 | -6.06 | Pal. 4 | 15.11 | -5.75 |
| N6355 | 1.72 | -7.36 | N5272 | 3.16 | -8.77 | N4372 | 5.22 | -7.48 |
| Lil. 1 | 1.34: | -7.42 | N5286 | 2.27 | -8.67 | Rup.106 | 6.50: | -6.29 |
| N6397 | 1.49: | -6.52 | AM4 | 3.48 | -1.50 | N4590 | 4.46 | -7.25 |
| Pal. 6 | 1.91: | -7.17 | N5634 | 3.86 | -7.64 | N4833 | 4.00 | -7.90 |
| Ter. 5 | 1.18 | -5.41 | N5694 | 3.17 | -7.70 | N5024 | 5.84 | -8.70 |
| N6440 | 1.27 | -8.57 | N5824 | 3.37 | -8.88 | N5053 | 16.29 | -6.64 |
| N6441 | 1.71 | -9.06 | N5927 | 2.34 | -7.66 | N5139 | 5.96 | -10.16 |
| Ter. 6 | 0.91 | -6.72 | N5946 | 2.39 | -7.47 | N5466 | 10.67 | -7.02 |
| N6453 | 1.13 | -6.93 | N5986 | 3.05 | -8.31 | I4499 | 7.68: | -7.21 |
| UKS 1 | 1.80: | -6.09 | N6121 | 2.23 | -7.06 | Pal. 5 | 18.77 | -5.04 |
| Ter. 9 | 1.68 | -3.78 | N6139 | 2.17 | -8.14 | N5897 | 7.61 | -7.18 |
| N6517 | 1.82 | -8.14 | N6205 | 2.95 | -8.43 | N5904 | 4.30: | -8.68 |
| N6535 | 1.48 | -4.62 | N6229 | 3.05 | -7.90 | Pal. 14 | 23.35 | -4.60 |
| N6528 | 0.88 | -6.53 | N6218 | 3.39 | -7.50 | N6101 | 7.31 | -6.80 |
| N6540 | 0.24: | -5.20 | N6235 | 2.30 | -6.01 | N6144 | 4.62 | -6.94 |
| N6544 | 1.24: | -6.44 | N6254 | 2.16 | -7.35 | Ter. 3 | 9.53: | -6.00 |
| N6553 | 1.85 | -7.59 | N6256 | 2.20 | -6.05 | N6171 | 4.71 | -6.98 |
| N6624 | 1.77 | -7.32 | N6266 | 2.36: | -9.11 | HP 1 | 7.30 | -7.60 |
| N6638 | 1.50 | -6.68 | N6273 | 3.02 | -8.97 | N6362 | 4.57 | -6.82 |
| N6637 | 1.88 | -7.35 | N6284 | 3.13 | -7.73 | Ter. 1 | 6.78 | -3.12 |
| N6642 | 1.55 | -6.44 | N6293 | 2.28 | -7.66 | N6401 | 4.00 | -7.47 |
| N6652 | 1.70: | -6.42 | N6304 | 2.30 | -7.15 | N6426 | 5.22 | -6.43 |
| Pal. 8 | 1.94 | -5.35 | N6316 | 2.25 | -8.46 | N6496 | 5.93 | -7.06 |
| N6717 | 1.31 | -5.48 | N6341 | 2.54 | -8.13 | N6558 | 4.07 | -6.73 |
| N6749 | 1.58: | -5.96 | N6325 | 2.46: | -7.20 | I1276 | 6.08 | -7.26 |
| N6838 | 1.73 | -5.40 | N6333 | 2.24 | -7.94 | N6760 | 4.38 | -7.72 |
| | | | N6342 | 2.20 | -6.37 | Ter. 7 | 6.18: | -4.88 |
| | | | N6356 | 2.97 | -8.35 | Pal. 10 | 4.69: | -5.48 |
| | | | N6352 | 3.08 | -6.31 | Arp 2 | 16.11 | -5.35 |
| | | | Ter. 2 | 3.94 | -4.86 | N6809 | 4.29 | -7.44 |
| | | | N6366 | 2.60 | -5.61 | Ter. 8 | 7.21: | -5.00 |
| | | | Ton. 2 | 2.58: | -5.16 | Pal. 11 | 5.16 | -6.63 |
| | | | N6388 | 2.12 | -9.65 | N6981 | 4.15 | -6.94 |
| | | | N6402 | 3.15 | -8.89 | N7006 | 4.37 | -7.58 |
| | | | Djo. 1 | 3.23: | -6.40 | Pal. 12 | 6.63 | -4.33 |
| | | | Djo. 2 | 3.19: | -8.80 | N7492 | 8.62 | -5.64 |
| | | | N6522 | 2.06 | -7.43 | | | |
| | | | N6539 | 3.64 | -8.13 | | | |
| | | | N6541 | 2.49 | -8.37 | | | |
| | | | N6569 | 3.13 | -7.72 | | | |
| | | | N6584 | 2.91 | -7.56 | | | |
| | | | N6626 | 2.50 | -8.20 | | | |
| | | | N6656 | 2.94 | -8.38 | | | |
| | | | N6681 | 2.33 | -7.03 | | | |
| | | | N6712 | 2.55 | -7.35 | | | |
| | | | N6715 | 3.62 | -9.89 | | | |
| | | | N6723 | 3.84 | -7.67 | | | |
| | | | N6752 | 2.59 | -7.62 | | | |
| | | | N6779 | 3.27 | -7.38 | | | |

## FIGURE CAPTIONS

Fig. 1  Plot of $M_V$ versus half-light radius $r_h$ for Galactic globular clusters. Clusters with 2 pc $< r_h <$ 4 pc are seen to be more luminous than both smaller and larger clusters. Metal-poor clusters ([Fe/H] $<$ -1.0) are shown as dots and metal-rich clusters ([Fe/H] $>$ -1.0) as open circles.

Fig. 2  Luminosity distribution of globular clusters binned into various size groups. Clusters with half-light radii of 2-4 pc are seen to be more luminous ($>$ 99% confidence) than both small clusters with $r_h <$ 2 pc <u>and</u> large clusters with half-light radii of 4-8 pc. Very large clusters with $r_h >$ 8 pc are found to be fainter than smaller clusters at 99.7% confidence.

Fig. 3  Integrated cluster magnitudes $M_V$ versus Galactocentric distance $R_{GC}$ for clusters with red horizontal branches having $C \equiv (B-R)/(B + V + R) <$ -0.80. The figure shows that outer halo clusters with red horizontal branches are faint.

Fig. 4  $M_V$ versus $R_{GC}$ for clusters with blue ($C >$ +0.80) and intermediate color (-0.80 $\leq C \leq$ +0.80) horizontal branches. Clusters with $C >$ +0.80 are shown as crosses, those with -0.80 ≤ C ≤ +0.80 are plotted as dots. The observed luminosity distributions of blue and intermediate-color clusters in the outer halo do not differ significantly from those with $R_{GC}$ < 10 kpc in the inner Galactic halo.

Fig. 5  $M_V$ versus [Fe/H] for clusters in the inner halo having $R_{GC}$ ≤ 10 kpc. No correlation is seen between $M_V$ and [Fe/H] over a range of ~100 in metallicity.

Fig. 6  $M_V$ versus [Fe/H] for halo globular clusters with $R_{GC}$ > 10 kpc. After excluding the anomalous metal-rich clusters Pal 1, Pal 12 and Ter 7, there is no correlation between metallicity and luminosity of clusters in the outer halo. The three anomalous relatively metal-rich and faint outer halo clusters may have had an unusual evolutionary history.

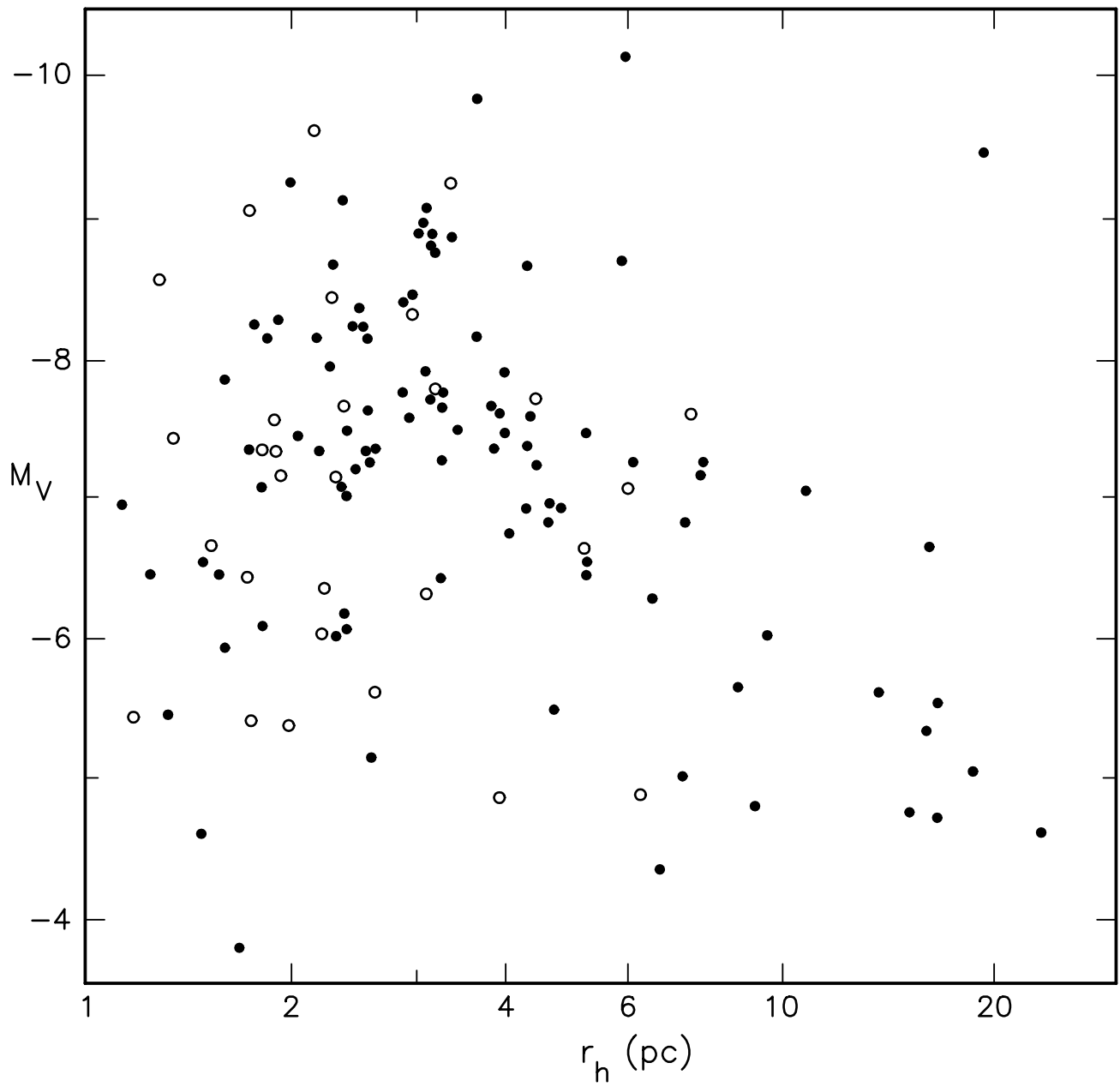

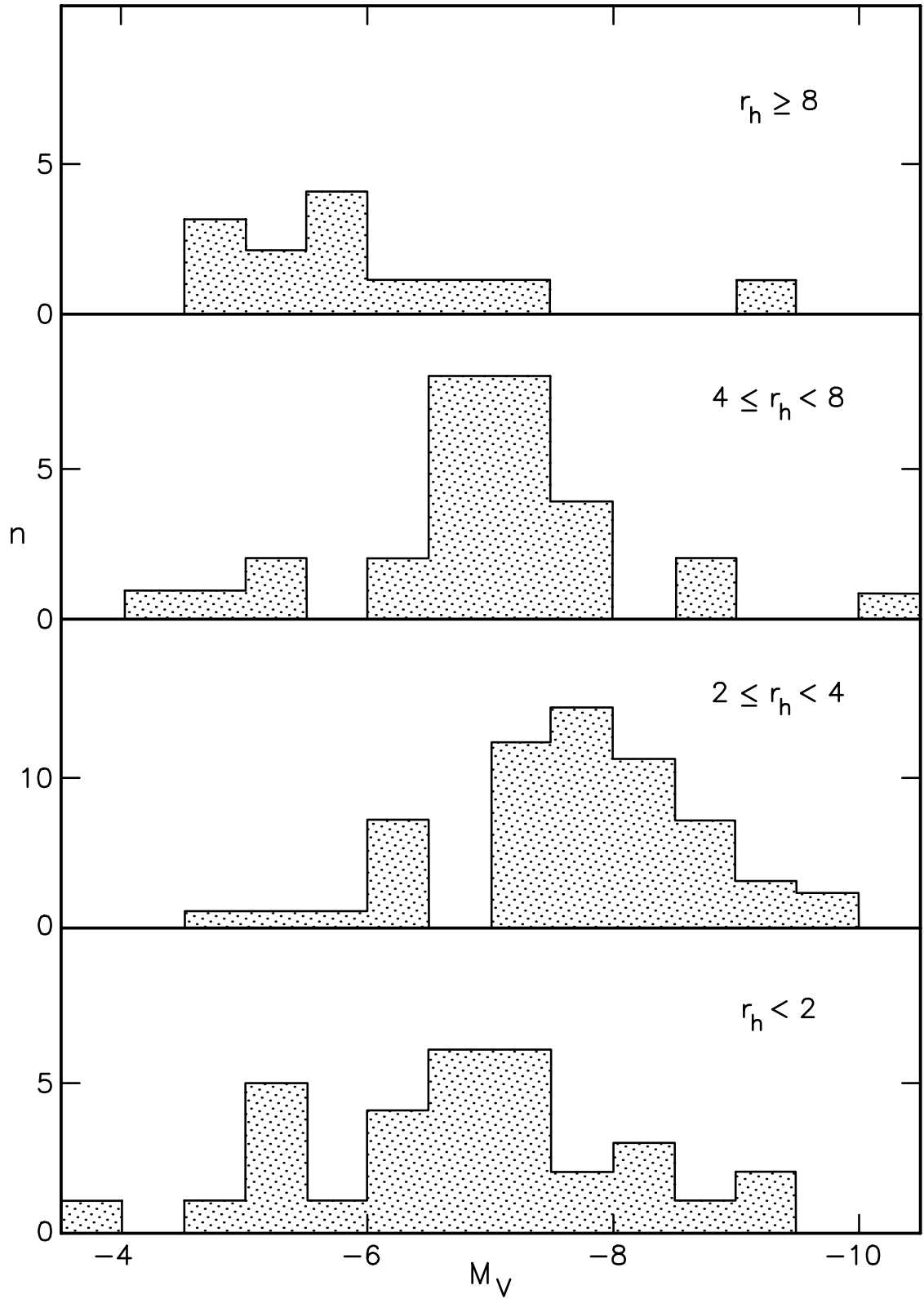

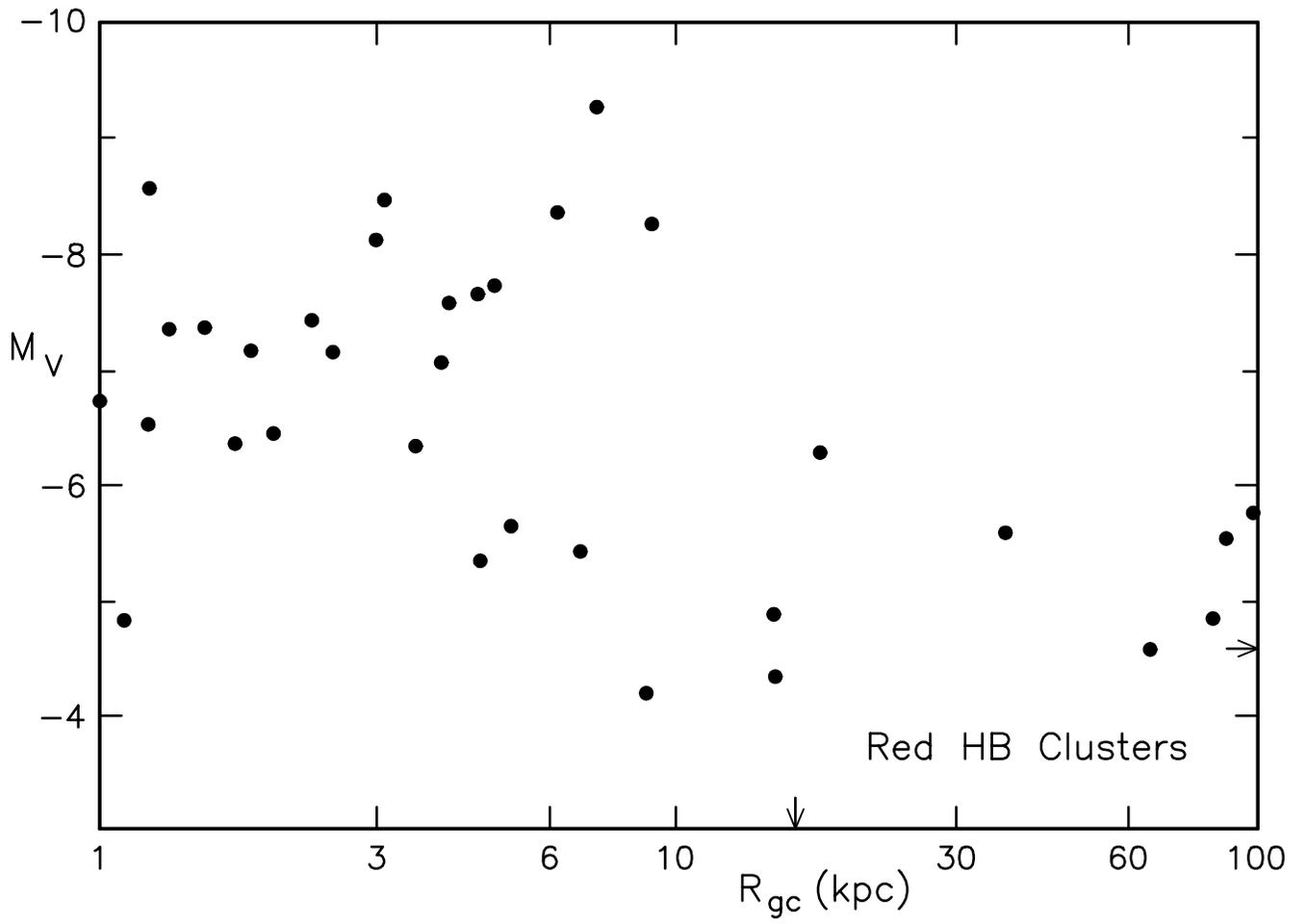

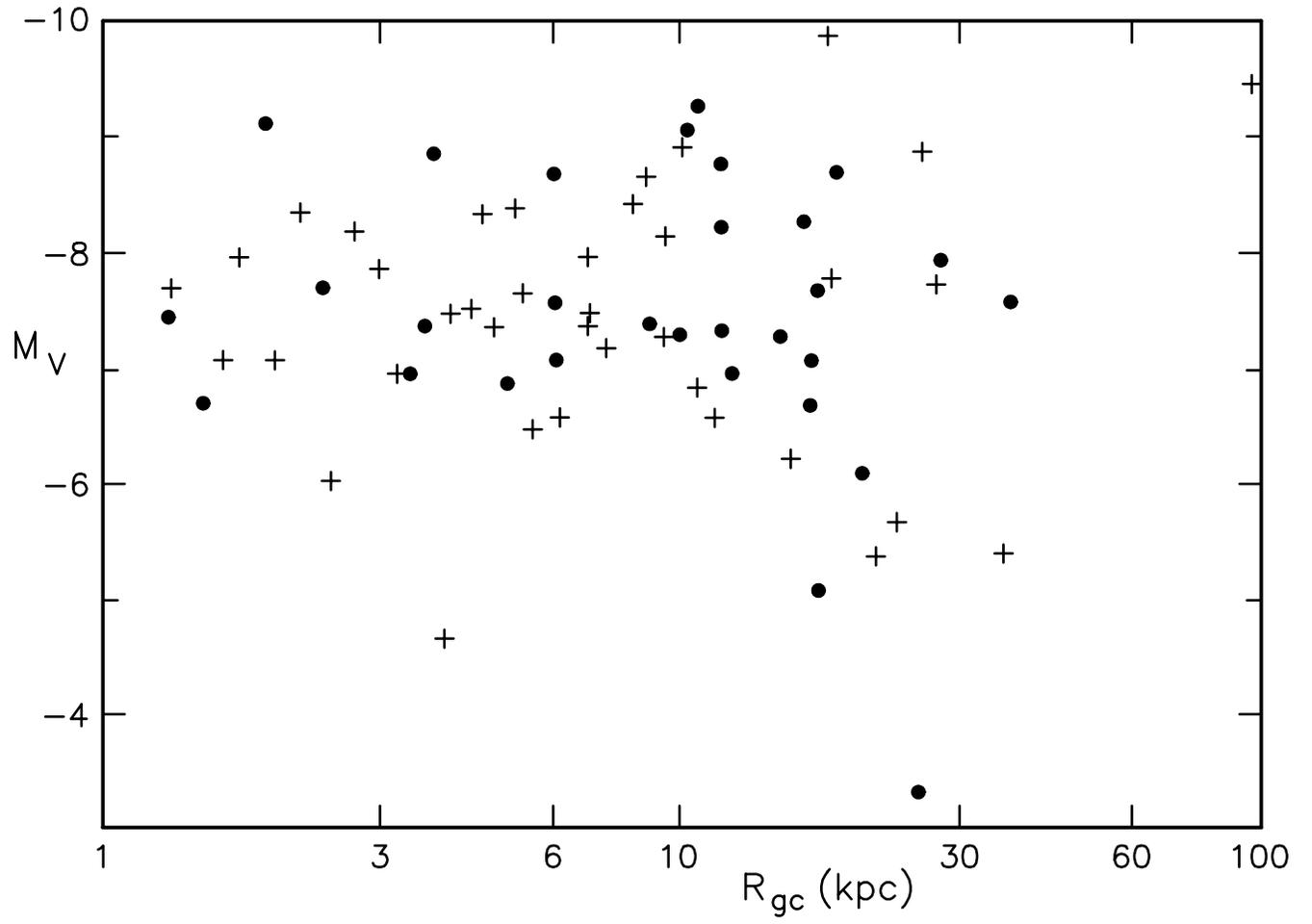

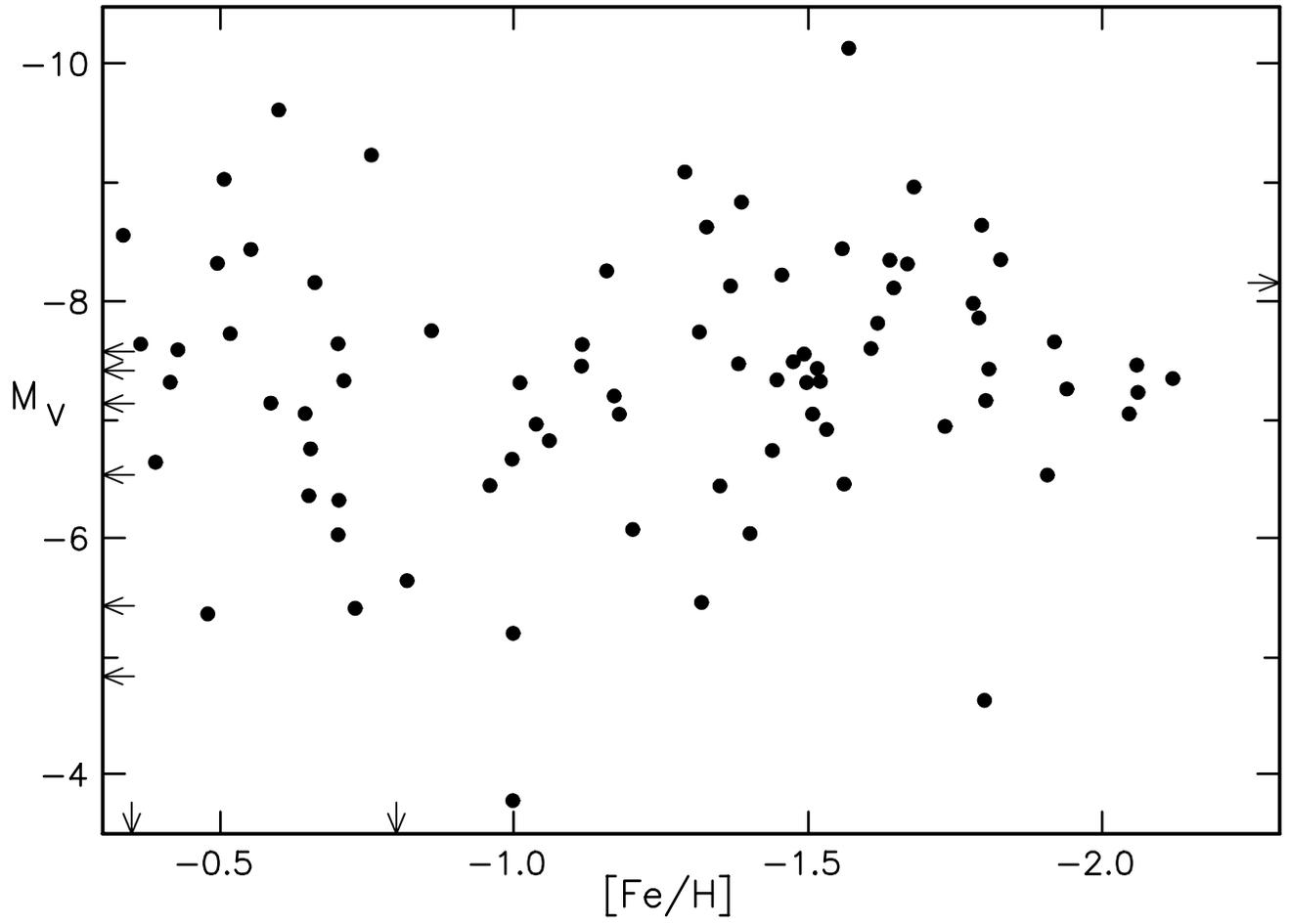

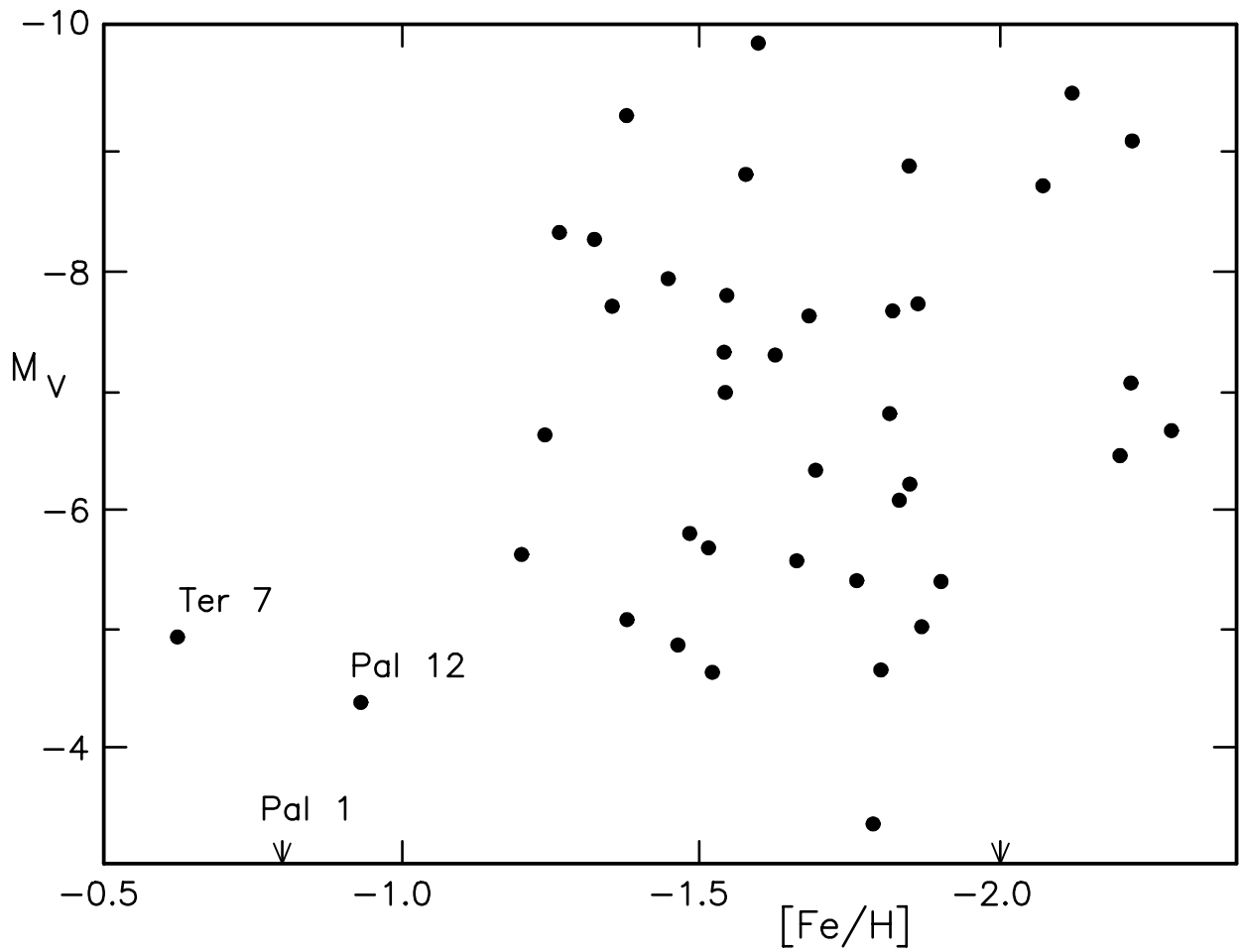